\documentclass[aps,twocolumn,floatfix,superscriptaddress]{revtex4}

\usepackage{amssymb,amsmath,amstext}                
\usepackage{graphicx}                                               
\usepackage{epstopdf}                                               
\usepackage{color}                                                     
\usepackage{bm}                                                        
\usepackage{appendix}                                              
\usepackage{bbold}
\usepackage{bbm}
\usepackage{ulem}
\usepackage{latexsym}
\usepackage[colorlinks=true,citecolor=blue,linkcolor=magenta]{hyperref}
\usepackage{cancel}
\usepackage{relsize}

\usepackage{soul}
\usepackage{mathtools}

\usepackage{pdfpages}

\newcommand{\vect}[1]{\mathbf{#1}}

\def\be{\begin{equation}}
\def\ee{\end{equation}}
\def\bea{\begin{eqnarray}}
\def\eea{\end{eqnarray}}
\def\ra{\rangle}
\def\la{\langle}
\def\bi{\begin{itemize}}
\def\ei{\end{itemize}}

\definecolor{dgreen} {RGB}{78,138,21}

\begin{document} 

\title{Vortex loop dynamics and dynamical quantum phase transitions in 3D fermion matter}

\author{Arkadiusz Kosior} 
\affiliation{Institute for Theoretical Physics, University of Innsbruck, 6020 Innsbruck, Austria}
\author{Markus Heyl} 
\affiliation{
Theoretical Physics III, Center for Electronic Correlations and Magnetism,
Institute of Physics, University of Augsburg, 86135 Augsburg, Germany}
\affiliation{Max-Planck-Institut f\"ur Physik Komplexer Systeme,
N\"othnitzer Strasse 38, D-01187, Dresden, Germany}

\date{\today}

\begin{abstract}

Over the past decade, dynamical quantum phase transitions (DQPTs) have emerged as a paradigm shift in understanding nonequilibrium quantum many-body systems. However, the challenge lies in identifying order parameters that effectively characterize the associated dynamic phases. 
In this study, we investigate the behavior of vortex singularities in the phase of the Green's function for a broad class of fermion lattice models in three dimensions after an instantaneous quench in both interacting and non-interacting systems. 
We find that the full set of vortices form one-dimensional dynamical objects, which we call \emph{vortex loops}. We propose that the number of such vortex loops can be interpreted as a quantized order parameter that distinguishes between different non-equilibrium phases. Our results establish an explicit link between variations in the order parameter and DQPTs in the non-interacting scenario. Moreover, we show that the vortex loops are robust in the weakly interacting case, even though there is no direct relation between the Loschmidt amplitude and the Green's function.
Finally, we observe that vortex loops can form complex dynamical patterns in  momentum space. 
Our findings provide valuable insights for developing definitions of dynamical order parameters in non-equilibrium systems.
\end{abstract}

\maketitle

\emph{Introduction.} Due to advancements in experimental technologies and a deeper theoretical understanding, the field of non-equilibrium quantum many-body dynamics has rapidly progressed in the recent decade~\cite{Polkovnikov2011}. This has led to the exploration of exciting, 
ergodicity-broken states and phases of quantum matter \cite{Mori2018,Moudgalya2022}, such as many-body scars \cite{Abanin2019} and time crystals \cite{Khemani2019,Sacha2020}, which are of inherent non-equilibrium nature and go beyond standard thermodynamics and the theory of quantum phase transitions \cite{Sachdev2000,Dziarmaga2010}. It has also been discovered that a~non-equilibrium unitary evolution of quantum states can give rise to temporal non-analyticities in the rate of the Loschmidt amplitude, which is analogous to the singularities observed in thermodynamic functions during equilibrium phase transitions.  This phenomenon, termed a dynamical quantum phase transition (DQPT) 
\cite{Heyl2013,Zvyagin2016,Heyl2018,Heyl2019}, has recently attracted a lot of attention from both theoretical \cite{Budich2016,Halimeh2017,Zunkovic2018,Lang2018,Heyl2018b,Kosior2018a,Kosior2018b,Yang2019,Zhou2021,Peotta2021,Trapin2021,Bandyopadhyay2021,Halimeh2021,DeNicola2021,Wong2022,Naji2022,Naji2022b,Jafari2022,Mondal2023,Cao2023,Wong2023} and experimental 
\cite{Flschner2017,Zhang2017,Guo2019,Tian2020,Jurcevic2017} communities (for related non-equilibrium transitions characterized by order parameters see  Refs.~\cite{Yuzbashyan2006,Sciolla2010,Halimeh2017b,Marino2022}). While it has been demonstrated in different cases that DQPTs are directly linked to the equilibrium phase transition of the system being studied, this connection cannot be considered a one-to-one correspondence in general \cite{Vajna2014}. Therefore, DQPT is expected to be a genuine non-equilibrium phenomenon without equilibrium counterpart, requiring the identification of order parameters that precisely characterize the dynamic phases in question. Remarkably,  DQPTs for a non-interacting system in two dimensions (2D) can be closely related to the emergence of measurable, dynamical vortex-like singularities in the phase of the Green's function in momentum space \cite{Yu2017,Qiu2018,Lahiri2019,Sim2022} (for experiments see \cite{Flschner2017,Tarnowski2019}). 
It has been shown so far that the vortices appear as stable objects in 2D two-band models, but higher-dimensional systems
have remained unexplored so far. 

In the following, we demonstrate that in a broad class of three-dimensional (3D) lattice models, with both interacting and non-interacting fermions, the vortices of the Green's function are not isolated;
instead, they form one-dimensional (1D) dynamical \emph{vortex loops}. We argue that the number of these vortex loops can be interpreted as a quantized order parameter that distinguish between different non-equilibrium phases.  Although focusing on a simple Weyl semimetal model with two bands \cite{Weyl1929,Delplace2012,Xu2015,Chen2015,Armitage2018,Wang2021}, we provide arguments and examples that our results are applicable to a wide class of fermionic translationally symmetric lattice models in 3D \cite{footExamples}. Importantly, we show that the vortex loops are robust and survive in weakly interacting systems, even though there is no direct relation between the Loschmidt amplitude and the Green's function.   Additionally, due to the existence of band touching Weyl nodes, we find that in a long time limit the loops can form complex dynamical patterns in  momentum space. 

\emph{General setup and observables.} 
Before we turn to a specific microscopic model, we start this section with a very  general picture. We consider a particle-hole symmetric model of non-interacting spin-$1/2$  fermions with translational invariance on a 3D lattice. The Hamiltonian of such a system can be written in a form:
\be\label{ham}
\hat H= \sum\nolimits_{\vect k\in \textrm{BZ}} \hat\psi_{\vect k}^\dagger H_{\vect k} \hat \psi_{\vect k},\quad H_{\vect k} = \vec{h}_{\vect k} \cdot \vec{\sigma},
\ee
with the summation going over independent quasi-momenta~$\vect k$ of a Brillouin zone (BZ). Here $\hat \psi_{\vect k}$ denotes a~fermionic spinor operator,
$\vec \sigma$ is a vector of Pauli matrices and $\vec{h}_{\vect k}$ is a vector
defined by microscopic details of a particular model.  Here, without a loss of generality, we consider   $\hat \psi_{\vect k}^\dagger = [\hat c_{{\vect k},\uparrow}^\dagger ,\hat c_{{\vect k},\downarrow}^\dagger  ]$ with  $ \hat c_{{\vect k},\sigma={\uparrow,\downarrow}}^\dagger$ being standard creation operators \cite{footN}, but we note that
 the precise form of $\hat \psi_{\vect k}$ might also be different depending on the choice of a two band model.  Throughout this work we adopt a unit lattice spacing and  set $\hbar=1$.
 
Furthermore, we assume that the system is prepared in a fermionic ground state at a half filling, namely,
$
|\psi_0 \ra = \Pi_{\vect k\in \mathrm{BZ}}|\vect k_-^{\mathrm{(i)}}\ra
$,
where $|\vect k_-^{\mathrm{(i)}}\ra = \hat c_{\vect k}^{\mathrm(i)\dagger} |0\rangle$  is the lower state of some initial Hamiltonian $H_{\vect k}^{\mathrm{(i)}} = \vec{h}^{\mathrm{(i)}}_{\vect k} \cdot \vec{\sigma}$, see~\cite{footE}.
At a time $t=0$ we perform an instantaneous quench of at least one of the model's  parameters, which results in a sudden change of the Hamiltonian $\vec{h}^{\mathrm{(i)}}_{\vect k} \rightarrow \vec{h}^{(\mathrm{f})}_{\vect k}$. After the quench, the initial state $|\psi_0\ra $ evolves 
 under the new Hamiltonian $H_{\vect k}^{(\mathrm{f})} = \vec{h}^{(\mathrm{f})}_{\vect k} \cdot \vec{\sigma} $,  i.e., $| \psi (t)\ra = \Pi_{{\vect k}\in \mathrm{BZ}} |\vect k(t)\ra$ with
\be\label{psi_t} 
 |{\vect k}(t)\ra 
 = \sum_{\alpha=\pm}\Gamma_{\vect k}^\alpha e^{-i  \epsilon_{\vect k,\alpha} t} |\vect k_\alpha^{(\mathrm{f})}\ra,\quad 
 \Gamma_{\vect k}^\alpha = \la \vect k_\alpha^{(\mathrm{f})}|\vect k_-^{\mathrm{(i)}} \ra,
\ee
where $| \vect k_\alpha^{(\mathrm{f})}\ra = \hat c_{\vect k,\alpha}^{\textrm{(f)}\dagger} | 0 \rangle $ is an eigenvector of 
$H_{\vect k}^{(\mathrm{f})}$ with the corresponding eigenenergy 
$\epsilon_{\vect k,\alpha=\pm}  = \pm \epsilon_{\vect k} $. 
We assume that the quench preserves translational invariance so that a quasi-momentum ${\vect k}$ remains a good quantum number. 

To investigate the dynamics of the non-equilibrium system we focus on the time ordered Green's function \cite{Peskin2018}
\be\label{green} 
g_{\vect{k}}(t) 
= \sum_{\vect r,\sigma} e^{i \vect{k}\cdot \vect{r}} \la \hat c_{\vect r,\sigma}(t)^\dagger \hat c_{\vect{r} = 0,\sigma}  \ra 
= \sum_{\vect q, \sigma}\la  \hat c^\dagger_{\vect k,\sigma}(t) \hat c_{\vect q,\sigma} \ra,
\ee
where  in the above we denote $\la \ldots \ra = \la \psi_0 | \ldots | \psi_0 \ra $ and $\hat c_{\vect r,\sigma} = 1/\sqrt{N} \sum_\vect{k} e^{-i \vect{k}\cdot \vect{r}}\hat c_{\vect k,\sigma} $. Alternatively,
one could also study the Loschmidt amplitude $\mathcal  G (t) = \langle \psi_0 | \psi (t) \rangle = \langle \psi_0 |e^{- i \hat H^{\textrm{(f})}t} | \psi_0 \rangle  $, which quantifies how far the time evolution drives the system away from the initial condition. The Loschmidt amplitude can be conveniently written as $\mathcal  G  (t)  =   \Pi_{\vect k} \mathcal G_{\vect k}^{(1)}(t)$, with 
\be
\mathcal G_{\vect k}^{(1)}(t) = \la 0 | \hat c_{\vect k}^{\textrm{(i)}\dagger}(t) \hat c_{\vect k}^\textrm{(i)}    | 0 \ra  = \la  \vect k_-^{\textrm{(i)}} | \vect k(t) \ra  \label{g1} .
\ee
Since the Loschmidt amplitude $\mathcal G$ for a many body system is a fast decaying function with the increasing number of particles $N$, it is   convenient to define the rate function, $\lambda= -\lim_{N\rightarrow \infty} \ln |\mathcal G|^2/N$. The latter bears formal resemblance to the free energy density (with temperature replaced by time $t$) and, therefore, $\lambda$ might be  viewed as a non-equilibrium free energy analog. This analogy implies that the rate function $\lambda$ can show signatures of a phase transition having non-analytic points that appear dynamically in time giving rise to a \emph{dynamical quantum phase transition} (DQPT) \cite{Heyl2013,Zvyagin2016,Heyl2018,Heyl2019}. For the considered setup  the rate function can be expressed analytically, i.e.,
\be\label{rate}
\lambda(t)= 
 -\int \frac{\mathrm{d}\vect k}{(2\pi)^3}\ln\left[1+\left[ \big(\hat h_{\vect k}^{\mathrm (i)} \cdot \hat h_{\vect k}^{\mathrm (f)}\big)^2-1\right]\sin^2(\epsilon_{\vect k}t) \right],
\ee
with 
 the normalised vectors 
$\hat h_{\vect k } = \vec h_{\vect k }/(\vec h_{\vect k } \cdot \vec h_{\vect k })^{1/2}$. 

To ensure clarity, throughout the article we have chosen to focus on a single model  while still drawing general conclusions \cite{footExamples}. Specifically, we consider a two-band 3D Weyl semimetal Hamiltonian  \cite{footW}, given by 
\be\label{hkweyl}
\vec h_{\vect k}= \left[ \sin k_x, \sin k_y, m_z - \cos k_z \right],
\ee
and we assume a quench of the free parameter $m_z$ between two distinct topological phases: a normal insulator ($|m_z|>1$), and a Weyl semimetal ($|m_z|<1$) characterized by pairs of Weyl points and linear Dirac-like dispersion around them. As we will demonstrate, this quench induces a DQPT and leads to the appearance of 1D dynamical singularities in the phase of the Green's function.

\emph{Relating DQPTs with phase singularities.}  For a non-interacting system,
it can be  shown  that $\mathcal G_{\vect k}^{(1)}  = \overline{g_{\vect k}} $, i.e., the Green's function is a complex conjugate of the first order correlation function  \cite{footG}, which implies that a DQPT can be observed on a Green's function level. Indeed, the non-analytic points of the rate function $\lambda(t)$ can only occur if and only if $ g_{\vect k}(t) = |g_{\vect k}(t)|\exp[i \phi_{\vect k}(t)]$ vanishes for some $\vect k^* $ and $t^*$, implying a phase singularity of the Green's function $\phi_{\vect k}(t)$ \cite{Heyl2018}.  While in  1D systems the phase singularities can be only observed in a $k-t$ plane \cite{Budich2016,Zache2019}, in 2D models these singularities appear as isolated  dynamical point vortices with clockwise or anticlockwise phase winding  \cite{Yu2017,Qiu2018,Lahiri2019,Sim2022} (for experiments see Refs.~\cite{Flschner2017,Tarnowski2019}). On the other hand, in the following sections we argue that for a wide class of 3D models these vortices group together forming dynamical 1D objects, i.e., \emph{vortex loops}, which can be either contractible or incontractible. The number of these objects can be associated with dynamical order parameters which identify different non-equilibrium phases. In turn, a change of a dynamical order parameter is accompanied by a DQPT, i.e., a temporal singularity of the rate function $\lambda$.  Later in this Letter we show that the vortex loops also appear in interacting systems, although there is no strict relation between the Green's function $g_{\vect k}$ and the Loschmidt amplitude $\mathcal G$ anymore.

\begin{figure}[t!] 	            
\includegraphics[width=0.95\columnwidth]{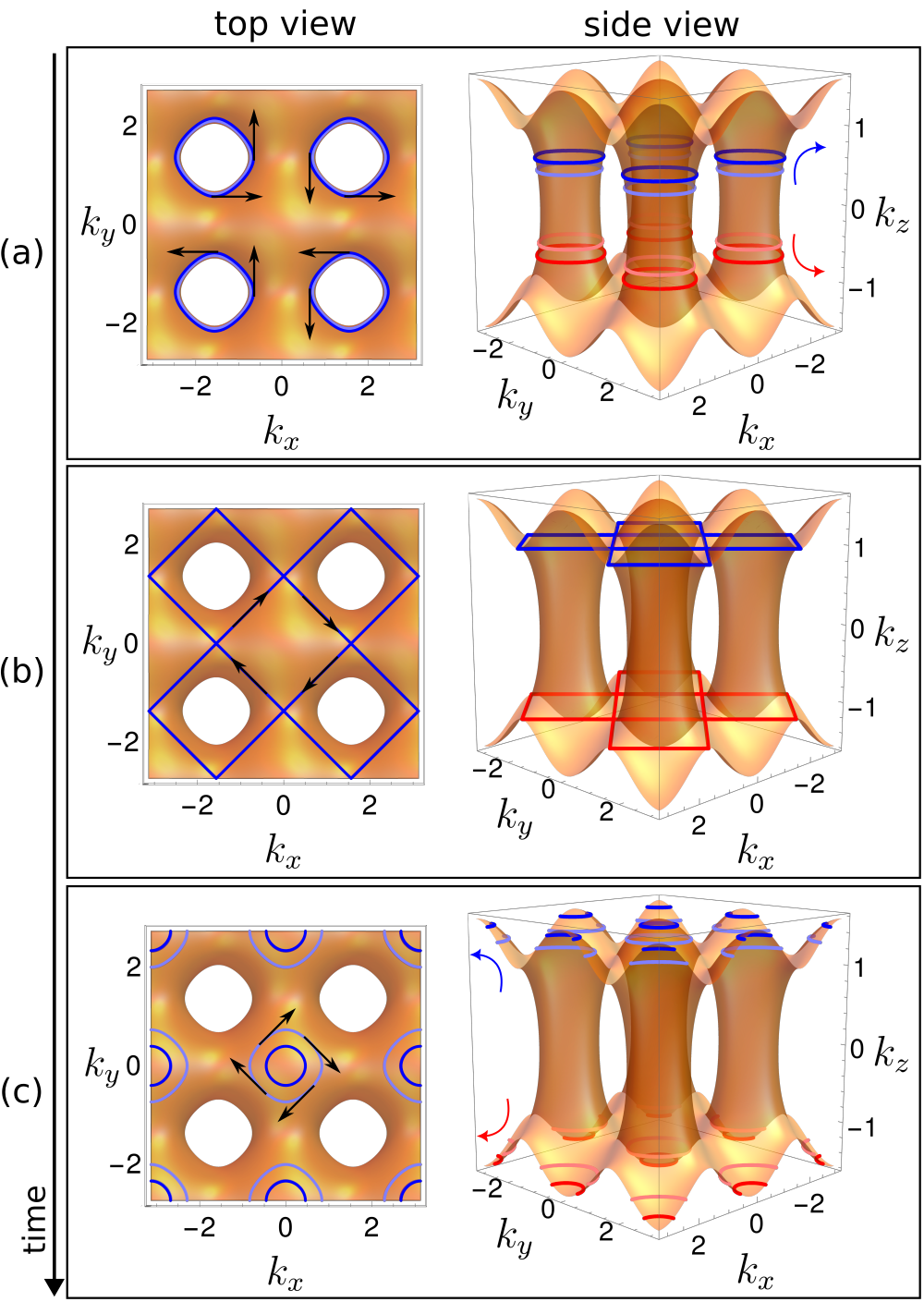} 
\caption{
Exemplary dynamics of vortex loops on a static manifold $\mathcal M$ (orange surface)
in a 3D BZ.  The red and blue colors represent the chirality of the loops. (a) Incontractible loop with opposing chiralities are created in pairs and move in opposite directions; light colors correspond to $t = 1.05$ and dark colors to $t=1.1$. 
(b) Loops of the same chirality can merge if their tangent vectors (black arrows) are antiparallel at the loops' touching points; $t= 1.405$ (c) Through loop merging, the loops can change their topological character from incontractible to contractible loops; $t = 2$ (light colors),  $t=3.5$ (dark colors).  
}\label{loops}
\label{dynamics}   
\end{figure}

\emph{Dynamics of vortex loops.}  For a generic, non-interacting two band model, a complex valued condition 
$
g_{\vect k}(t) = \sum_\alpha |\Gamma_{\vect k}^\alpha|^2  e^{i\epsilon_{\vect k,\alpha} t}  = 0
$  
implies
\be
\label{conds} 
(i) \; \mathcal{M}:
|\Gamma_{\vect k}^\pm|^2=\frac{1}{2},
\quad (ii) \; \mathcal{M}_n(t):
  t_n =   \frac{(2n+1)\pi}{2\epsilon_{\vect k}},
\ee
with  $n \in \mathbb{N}$, $\epsilon_{\vect k,\alpha=\pm}  = \pm \epsilon_{\vect k} $ and $\Gamma_{\vect k}^\alpha$ defined as in Eq.~\eqref{psi_t}. In 3D, the conditions in Eq.~\eqref{conds} define two 2D surfaces in the momentum space. The first condition defines a static manifold, $\mathcal M$, illustrated as the orange surface in Fig.~\ref{loops}. The second condition defines a dynamical, equienergy surface, $\mathcal{M}_n(t)$, that can intersect with $\mathcal M$ during some time intervals. 
Here we assume that $\mathcal{M}$ and $\mathcal{M}_n(t)$ are smooth 2D surfaces within the 3D Brillouin zone \cite{comment1}. The 3D Brillouin zone, being a 3-torus, is a compact manifold with no boundary, i.e., it is a closed manifold. Consequently, also $\mathcal{M}$, $\mathcal{M}_n(t)$ and their non-empty intersection is also closed and smooth, consisting of closed curves (i.e., the vortex loops) and/or isolated points when the vortex loops are being created or annihilated \cite{comment2}.

The vortex loops in different stages of evolution are illustrated in Fig.~\ref{loops} by the blue and red curves. For our analysis, we have selected a 3D Weyl Hamiltonian \cite{footW}, determined by Eq.~\eqref{hkweyl}, with $m_z^{(i)}=2.5$ and $ m_z^{(f)}=0$.  In this case,  $\mathcal M$ is a connected surface with a non-zero genus~\cite{Hirzebruch1966} and, therefore, any closed curve on this surface can be classified as either a contractible or incontractible loop. Further examples of $\mathcal M$ can be found in the Supplemental Material \cite{footExamples}.




\begin{figure}[t!] 	            
\includegraphics[width=0.9\columnwidth]{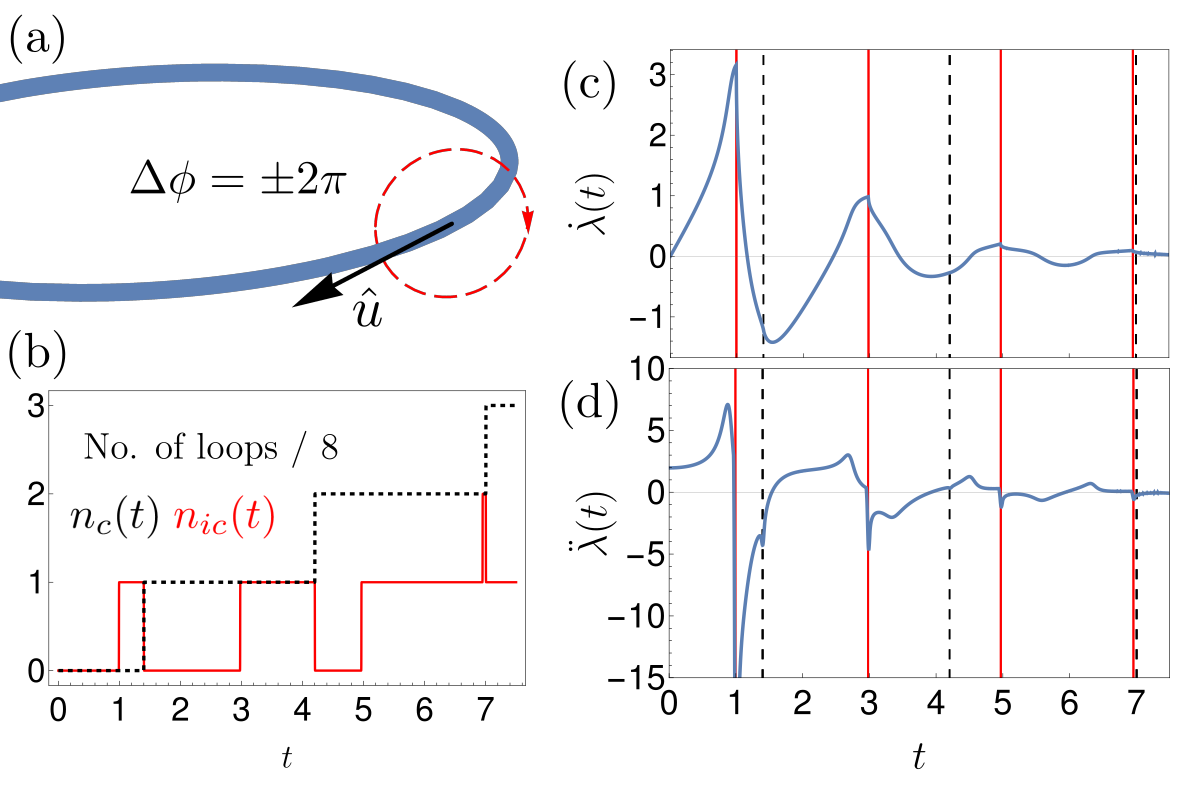} 
\caption{(a): Chirality ($\pm$) of a vortex loop is determined by $\Delta \phi = \pm 2\pi$ phase jump while encircling a loop around a tangent vector $\hat u$.  (b): Number of contractable  $n_c(t)$ and incontractible $n_{ic}(t)$  loops can be considered as quantized dynamical order parameters (black dashed and red solid lines, respectively).  
(c) and (d): The first and second derivative of the rate function $\lambda(t)$ with non-analytic DQPT points (marked by vertical lines),  exactly match the changes of  $n_c(t)$ and $n_{ic}(t)$.   }
\label{fig2}   
\end{figure}



In addition, the loops can also be characterized based on their chirality, indicated by $\pm$, see Fig.~\ref{fig2}(a). Specifically, any closed curve circulating a vortex loop undergoes a phase jump of $\Delta \phi = \pm 2\pi$, where the sign of the jump represents the chirality of the loop. Loops' chiralities are important in terms of their dynamics, e.g., loop merging or loop creation as explained in the following.

In a typical scenario, the vortex loops are created at certain times, $t_n^{\mathrm (c)}=(2n + 1)\pi/(2\epsilon_{\mathrm{max}})$, and annihilated at $t_n^{\mathrm (a)}=(2n + 1)\pi/(2\epsilon_{\mathrm{min}})$, $n\in\mathbb{N}$, where  $\epsilon_{\mathrm{max}}$ ($\epsilon_{\mathrm{min}}$) is the maximal (minimal) value of the upper band dispersion relation over quasi-momenta $\vect k$ belonging to $\mathcal M$, i.e., $\epsilon_{\mathrm{max}}=\max_{\vect k \in \mathcal M} \epsilon_{\vect k}$ ($\epsilon_{\mathrm{min}}=\min_{\vect k \in \mathcal M} \epsilon_{\vect k}$). As we show in Fig.~\ref{loops}~(a), the loops can be created (or annihiliated) in pairs with opposing chiralities, represented by red and blue colors. Alternatively, contractible loops can be also created from a single point of the BZ and annihilated in the inverse process.  In the course of time evolution, the vortex loops of the same chirality can merge together as long as their tangent vectors at the touching point are antiparallel, for example, see Fig.~\ref{loops}~(b). Through the loop merging, it is even possible that the loops change their topological character, from incontractible to contractible [cf.~Fig.~\ref{loops}~(c)], or vice versa. 

The dynamics of vortex loops is rather complex, but can be captured qualitatively and quantitatively by monitoring the number of loops over time, denoted as $n_c(t)$ and $n_{ic}(t)$ for contractible and incontractible loops, respectively (Fig.\ref{fig2}(b)). These numbers of vortices can be interpreted as quantized dynamical order parameters that distinguish between different non-equilibrium phases. To support this interpretation, we plot the first and second derivative of the rate function $\lambda(t)$, given by Eq.\eqref{rate}, in Fig.\ref{fig2}~(c)-(d). In this work, we study a parameter quench from a normal insulator phase to a Weyl semimetal phase \cite{footW}, which passes through a critical point of an equilibrium quantum phase transition, resulting in a change in the topological properties of the underlying Hamiltonian. Our analysis reveals a sequence of non-analytic points in time, which corresponds to a series of DQPT events, as per the definition \cite{Heyl2018}. Within our model, we observe two types of non-analytic points: those corresponding to a discontinuity in the second derivative of $\lambda(t)$  and those corresponding to a divergence of $\ddot \lambda(t)$. By comparing the critical times with the number of vortex loops of each kind, we observe that the appearance of each non-analytic point is necessarily associated with a change in $n_c(t)$ or $n_{ic}(t)$.

\begin{figure}            
\includegraphics[width=1\columnwidth]{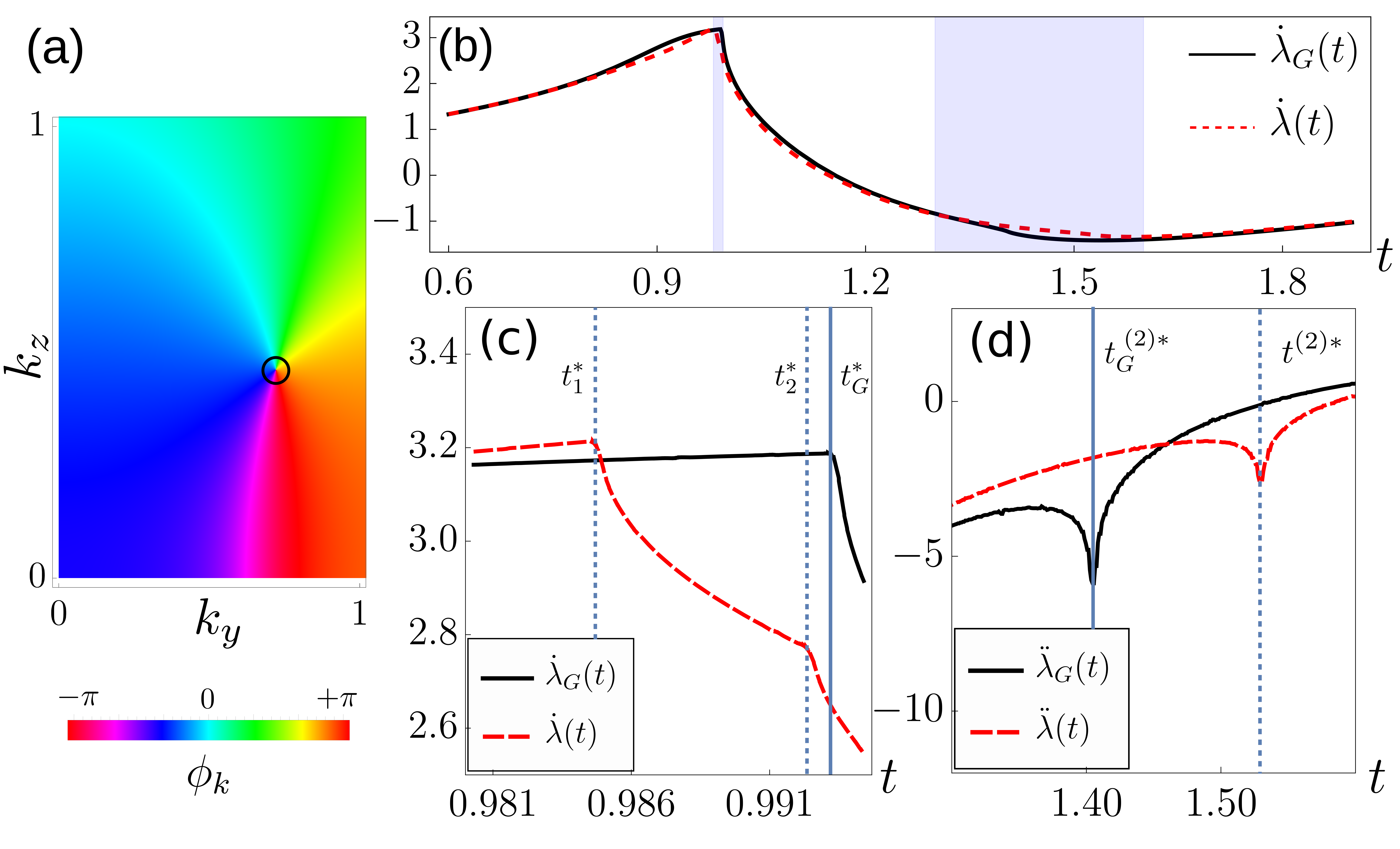} 
\caption{
(a) The phase of the Green's function $g_{\vect k}(t)$ in the interacting regime is shown for interaction strength $\eta=0.1$ and time $t=1.05$, with a fixed value of $k_x = \pi/2$. The black circle marks the position of the vortex in the non-interacting case. (b) The first derivatives of the rate function $\lambda(t)$ (red dashed line)  and $\lambda_G(t)$ (black solid), the later being defined in Eq.~(10). Although the two curves almost overlap, their non-analytic points do not coincide and their mismatch grow in time.  Non-analytic points of $\lambda(t)$ and $\lambda_G(t)$ are contained within the two shaded areas, as depicted in panels (c) and (d), respectively.  (c) 
Zoom on $\dot \lambda(t)$ and $\dot \lambda_G(t)$ in a vicinity of the first temporal non-analytic points at $t_1^*$, $t_2^*$ and $t_G^*$. (d) Zoom on $\ddot \lambda(t)$ and $\ddot \lambda_G(t)$ showing the second type of non-analytic points at $t^{(2)*}$ and $t^{(2)*}_G$. 
}\label{int}
\end{figure} 

\emph{Effects of interactions.} So far, we have focused on a non-interacting fermionic model. In this section we aim to demonstrate that our findings are more general and robust by showing that the vortex loops persist even in the presence of weak interactions. Consequently, let us consider   $H = H_0 + \eta V$, with $\eta$ being a small interaction strength and $V$ being  a generic two-body interaction term
To calculate the Green's function $g_{\vect k}(t)$ in the interacting regime, the only difficulty lies in obtaining the time evolution operator
in the interaction picture $U(t)$. However, up to linear order in $\eta$, we can utilize the Magnus expansion  \cite{Magnus1954,Blanes2009} and truncate  $V$ to its leading terms~\cite{Kriel2014} in order to approximate 
$U(t) \approx e^{-i H_0^f t} \exp\left(-\Omega \right)$  
with 
\be\label{evolution_op_in}
\Omega(t) \approx - i \eta t \sum_{\vect k}\sum_{\alpha,\beta} \Lambda_{\vect k,\vect k'}^{\alpha,\beta}\;
\hat n_{\vect k, \alpha}^{\textrm{(f)}} \hat n_{\vect{k'}, \beta}^{\textrm{(f)}},
\ee
where coefficients $\Lambda_{\vect k,\vect k'}^{\alpha,\beta}$ depend on the specific interaction type. Here, we choose BCS interactions and get $\Lambda_{\vect k,\vect k'}^{\alpha,\beta} = \delta_{\vect{k'},\vect{\text{-}k}}|\gamma_{\vect k, \uparrow}^\alpha |^2  |\gamma_{\vect{k'}, \downarrow}^\beta |^2 $ with~$\gamma_{\vect k, \sigma}^\alpha = $ $ \la \vect k_\alpha^{\textrm{(f)}} | \vect k_\sigma \ra$, see \cite{footI}. Inserting Eq.~\eqref{evolution_op_in} into the formula for the Green's function, Eq.~\eqref{green}, one readily gets
\be
g_{\vect k}(t) \approx
\sum_{\alpha,\beta} |\Gamma_{\vect k,\alpha}\Gamma_{\text{-}\vect k,\beta}|^2 e^{i t \epsilon_{\vect k, \alpha} } e^{-it \left(\Lambda_{\vect k,\text{-}\vect k}^{\alpha, \beta}  + \Lambda_{\text{-}\vect k,\vect k}^{ \beta, \alpha}\right)}
\ee
with $\Gamma_{\vect k}^\alpha = \la \vect k_\alpha^{(\mathrm{f})}|\vect k_-^{\mathrm{(i)}} \ra$. 
Following these steps, in Fig.~\ref{int}(a) we plot the phase of  $g_{\vect k}(t)$, choosing the interaction strength $\eta=0.1$ and a time $t=1.05$.  For clarity of the presentation we fix $k_x = \pi/2$  and illustrate a 2D cut through momentum space, which shows  a clear phase singularity close to the center of the panel.  Although the interacting part of the Hamiltonian   adds a correction to the Green's function, we find that within the numerical precision the position of the vortex in the momentum space matches the position of the vortex in the non-interacting case, marked by a black circle in Fig.~\ref{int}(a). 

As discussed, in the non-interacting system, the vortex loop dynamics is inherently imprinted in the rate function $\lambda(t)$. In the following we explore whether this relation still holds for the interacting case. In Fig.~\ref{int}(b) we illustrate the first derivative of the rate function $\lambda(t)$ (the red dashed line) in the vicinity of the first non-analytic point in time, which, for $\eta =0.1$ is close to $t^*\approx 0.985$. In order to quantitatively determine the point in time associated with the loops' creation and directly compare the behaviour of the Loschmidt amplitude and the Green's function, we define 
\be\label{lambdaG}
\lambda_G(t) = -\lim_{N\rightarrow \infty} \ln |\Pi_{\vect k} g_{\vect k}(t) |^2/N
\ee
and plot its first derivative in Fig.~\ref{int}(b) (the black solid line). Although on the first glance the first derivatives of $\lambda(t)$ and $\lambda_G(t)$ exhibit strikingly close overlap it is noteworthy that their non-analytic points do not coincide [Fig.~\ref{int} (c)-(d)]. On the contrary, we find that the discrepancy between the non-analytic points grows in time.

\begin{figure}[t!] 	    
\centering
\includegraphics[width=0.9\columnwidth]{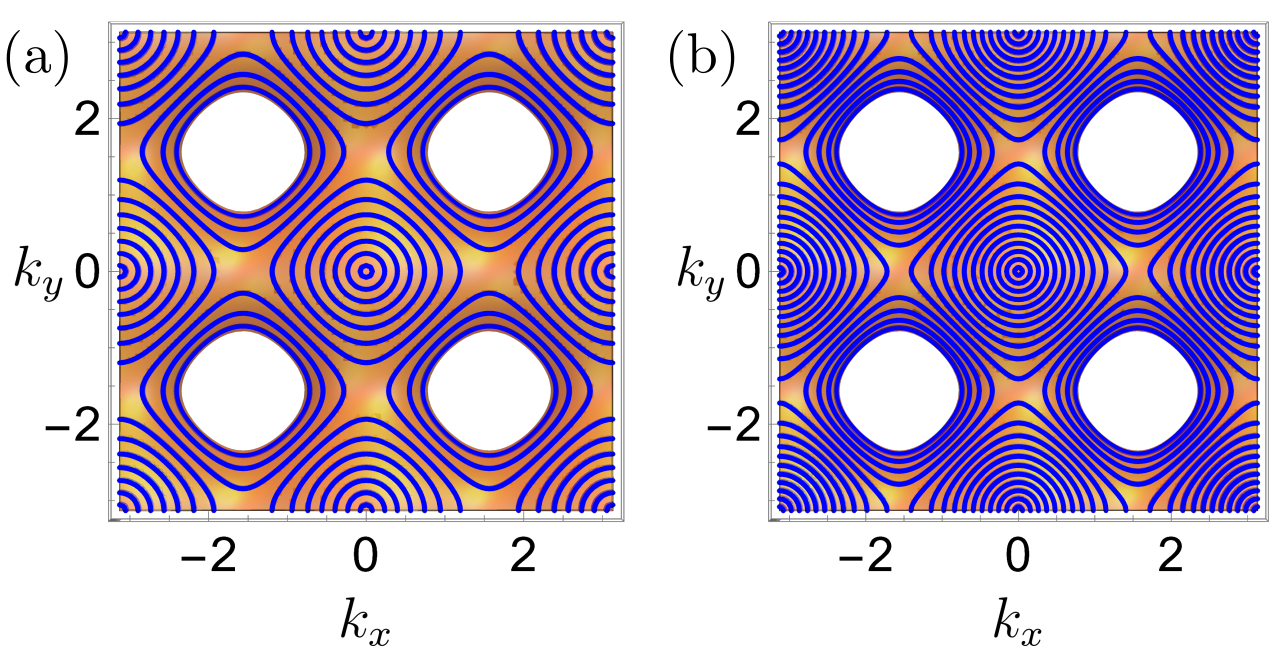} 
\caption{The complex pattern formed by loop vortices at times $t=15$ [panel (a)] and $t=30$ [panel (b)] after the quench to the Weyl semimetal phase, where the number of loops around the Weyl nodes increases linearly in time. Due to the dispersion relation $\epsilon_{\vect k}$ vanishing for some points belonging to the manifold $\mathcal M$, the loops never annihilate, and instead their dynamics slows down, leading to their accumulation around the band touching points $\vect k_W$. The resulting dynamical pattern cannot be destroyed due to the system's dynamics.}
\label{pattern}   
\end{figure} 

\emph{Dynamical pattern formation.} 
Although the loop dynamics is rather complex, the loop creation and annihilation  time can be determined easily from Eq.~\eqref{conds}. In particular, the annihilation time is given by $t_n^{\mathrm{(a)}}=(2n+1)\pi/(2\epsilon_{\mathrm{min}})$, $n\in\mathbb{N}$, where $\epsilon_{\mathrm{min}}=\min_{\vect k\in \mathcal M} \epsilon_{\vect k}$.
However, in our analysis presented in Fig.\ref{fig2}(b), the total number of loops never decreases in time due to the choice of the model and the quench to the Weyl semimetal phase, which hosts pairs of the band touching Weyl nodes denoted by $\vect k_W$ \cite{footW}. As the dispersion relation $\epsilon_{\vect k}$ vanishes for some points belonging to the manifold $\mathcal M$, the loops are never annihilated. Instead, the loops' dynamics slow down, and eventually, they accumulate around the band touching points $\vect k_W$. In Fig.~\ref{pattern}, we show a complex pattern formed by the loop vortices at times $t=15$ and $t=30$ after the quench. Although the number of loops around the Weyl nodes increases linearly in time, the resulting dynamical pattern cannot be destroyed due to dynamics. 

\emph{Summary and perspectives.}
The Green's function plays a crucial role in quantum many-body theory. Our  main finding is that the Green's function in 3D fermion matter after a parameter quench involves the dynamical creation and annihilation of topological defects in momentum space in the form 
of vortex loops. We have demonstrated that these loops are triggered by a underlying dynamic quantum phase transition. As a consequence, these vortex loops act as a dynamic topological
order parameter.
Moreover, we have shown that the vortex loops survive in weakly interacting systems, and that  they can  form complex dynamical patterns in momentum space due to the existence of band touching points. 

Our findings reveal that non-equilibrium dynamics in 3D systems exhibit a significantly greater level of complexity compared to lower dimensions. As a result, it appears that to obtain a deeper understanding of the intricate out-of-equilibrium quantum phases of matter, there is a need to undertake further investigations on DQPTs specifically in 3D systems. We emphasize the importance of exploring further the connection between DQPT and vortex loops in interacting systems, especially beyond perturbative regimes. Such exploration could potentially prompt a reevaluation of DQPT, shifting the focus from the Loschmidt amplitude to the involvement of the Green's function.

\begin{acknowledgments}
This research was funded in whole or in part by the Austrian Science Fund (FWF) [10.55776/ESP171]. This project has received funding from the European Research Council (ERC)
under the European Union’s Horizon 2020 research and innovation programme
(grant agreement No. 853443).  
\end{acknowledgments}


\begin{thebibliography}{68}
\bibitem[{com({\natexlab{a}})}]{comment1}
\bibinfo{note}{The assumption that $\mathcal{M}$ are $\mathcal{M}_n(t)$ are 2D
  smooth surfaces is well justified as long as the Hamiltonian before and after
  the quench is described by smooth functions $\vec h_{\vect k}^{\mathrm (i)} $
  and $ \vec h_{\vect k}^{\mathrm (f)}$, which are also bounded as the lattice
  Hamiltonian $H_{\vect k}$ has a bounded spectrum, both from below and above.}
\bibitem[{com({\natexlab{b}})}]{comment2}
\bibinfo{note}{This assumption is made under the condition that there are no
  surface parts touching, which represents a finely-tuned scenario, for example
  where $\mathcal{M}: f_1(k_x) = 0 $ and $\epsilon_{\vect k} = f_2(k_x,t)$
  which reduces to the 1D problem.}
\end{thebibliography}


\end{document}